\documentstyle[prl,aps,manuscript]{revtex}

\font\bba=msbm10 scaled 1200
\font\bbb=msbm8 %scaled 1080
\font\bbc=msbm6 %scaled 1080
\newfam\bbfam
\textfont\bbfam=\bba
\scriptfont\bbfam=\bbb
\scriptscriptfont\bbfam=\bbc

\begin{document}
\draft    
\title
{Transport Coefficients of the Yukawa One Component Plasma.}
\author{Gwena\"el Salin \thanks{e-mail:
salin@labomath.univ-orleans.fr}}
\address{MAPMO - CNRS (UMR 6628)\\
D\'epartement de Math\'ematiques \\
         Universit\'e d'Orl\'eans, BP 6759\\
         45067 Orl\'eans cedex 2, France}
\author{Jean-Michel Caillol \thanks{e-mail: caillol@lyre.th.u-psud.fr}}
\address{LPT - CNRS (UMR 8627) \\
	 B\^atiment 210,
	 Universit\'e Paris Sud \\
	 F-91405 Orsay Cedex, France}
\date{\today}
\maketitle
\begin{abstract}
    We present  equilibrium molecular-dynamics computations of 
the thermal conductivity and the two viscosities  of the 
Yukawa one-component plasma. The simulations were performed within
periodic boundary conditions and Ewald sums were implemented for the potentials, 
the forces, and for all the currents which enter the Kubo formulas. 
For large values of
the screening parameter, our estimates of the shear viscosity and the 
thermal conductivity are in good 
agreement  with the predictions of the Chapman-Enskog theory. 
\end{abstract}

\narrowtext
Recently, many numerical studies of the Yukawa one-component plasma 
(YOCP)
- i.e. a system made of $N$ identical classical point particles of charge
$q$ and mass $m$ which 
are embedded 
in a uniform neutralizing background of volume $V$ and which interact 
via Yukawa pair potentials $v(r)=q^{2}\exp (-\alpha r)/r$ -
have been performed in view of 
applications for a broad variety of systems, including dusty plasmas, 
inertial confinement fusion dense  plasmas, jovian planets, brown and 
white dwarfs, etc. The excess free energy $f$ as well as all the
thermodynamic properties  of the YOCP depend only upon two parameters, 
namely the coupling parameter $\Gamma= \beta q^{2}/a$, where 
$\beta=1/kT$ is the inverse temperature and $a$ the ionic radius ($4 
\pi \rho a^{3}/3=1$,  $\rho = N/V$ number density), and 
the reduced screening parameter $\alpha^{*}=\alpha a$. In the special 
case where $\alpha^{*}=0$ one recovers the well-known one-component 
plasma (OCP) 
\cite{Hansen}. The other limiting case  $\alpha^{*} \to \infty$ is that of a 
dilute gase for which simple approximate schemes can safely be used.
The thermodynamic and structural properties of the YOCP have been 
thoroughly studied  by means of equilibrium  molecular dynamics (EMD) 
simulations within periodic boundary conditions (PBC) \cite{Hama1,Hama2} 
and by Monte Carlo simulations on the hypersphere\cite{CG}.
Reliable estimates 
of the free energy $f(\Gamma,\alpha^{*})$ are thus  available in a 
wide range of $(\Gamma,\alpha^{*})$\cite{Hama1,Hama2,CG}. 

By contrast, very 
little is known about the dynamical properties of the model.
In view of hydrodynamical simulations, precise estimates of the transport 
coefficients of the YOCP are clearly wanted.  Attempts to compute the
shear viscosity $\eta$  by means of 
non-equilibrium molecular dynamics (NEMD) simulations  
were discussed recently in the literature \cite{Murillo}.  
In this letter we present (EMD)
computations not only of $\eta$,
but also of the bulk 
viscosity $\xi$ and the thermal conductivity  $\lambda$. It turns out 
that our results for $\eta$ differ significantly from those of 
ref.\cite{Murillo}, a puzzling point which will be discussed further on.

As it is well known, the three
transport coefficients $\eta$, $\xi$, and  $\lambda$ are given by  
 the Kubo formulas \cite{Hansen2,Til,Bernu} : 
\begin{mathletters}
\label{Kubo}    
\begin{equation}
\eta  =  \frac{\beta}{V} \int_{0}^{\infty} \langle \sigma_{xy}(t)
\sigma_{xy}(0) 
\rangle \;  dt \; , 
\end{equation}
\begin{equation}
\xi  =  \frac{\beta}{9 V}  \sum_{\alpha,\beta} \int_{0}^{\infty} 
\langle \sigma_{\alpha \alpha}(t)
\sigma_{\beta \beta}(0) \rangle \;  dt  \; ,
\end{equation}
\begin{equation}
    \lambda= \frac{1}{3 V kT^{2}} 
     \int_{0}^{\infty} \langle \vec{J}_{e}(t) \cdot \vec{J}_{e}(0) \rangle \; 
dt \; .
\end{equation}    
\end{mathletters}
In Eqs.\ (\ref{Kubo})  $\sigma_{\alpha \beta}$ denotes the Fourier 
transform
of one of the cartesian components of the pressure tensor at 
$\vec{k}=\vec{0}$ and  $\vec{J}_{e}$ is the
$\vec{k}=\vec{0}$ component of the Fourier transform of the energy current. 

 Our simulations were performed in a cube of side $L$ with PBC 
conditions and we 
took an explicit account of the periodicity of the system by making use 
of Ewald sums. We have shown elsewhere that the PBC expression of the 
Yukawa pair potential reads, up to an additional constant, as \cite{SC}:
\widetext
\begin{equation}
    \label{Ewald}
v_{PBC}(\vec{r}) = \frac{4\pi q^2}{L^{3}} \;\sum_{ \|\vec{k}\|
\le k_0} \frac{\exp\left(-\left(\vec{k}^2+\alpha^{2}\right)/4 \delta^{2}
\right)}
{\vec{k}^2+\alpha^{2}} \exp(i\vec{k} \cdot \vec{r}) 
+ q^2 \sum_{\epsilon=\pm 1}\frac{\text{erfc}(\delta \|\vec{r}\| +\epsilon 
\alpha /2 \delta)\; \exp( \epsilon \alpha \| \vec{r} \| )}{2 \|\vec{r}\|}\; ,   
\end{equation}  
\narrowtext   
where  the sum in the r.h.s runs over the vectors 
$\vec{k}$  of the reciprocal lattice. The 
parameter $\delta$ is  chosen in such a way that the 
contributions to     $v_{PBC}(\vec{r})$  in the direct space
reduce to a single term ( i.e. the 
second term of the r.h.s. of Eq.\ (\ref{Ewald})) and that  the cut-off 
$ k_0 $ on the vectors $\vec{k}$ is not too large. The optimal choice,
which ensures a 
relative precision of the order of $\sim 10^{-6}$ on  $ 
v_{PBC}(\vec{r})$ for all the points $\vec{r}$ inside  the simulation cell
is $\delta \times L \sim 5.6$ \cite{SC}. 

The Ewald expressions for the pressure tensor $\sigma_{\alpha \beta}$ 
and the energy current $\vec{J_{e}}$ can be  obtained by generalizing the 
pioneer work of Bernu and Vieillefosse on the OCP \cite{Bernu}. The 
details of the derivation will be given elsewhere, and we just quote 
here, as an example, 
the  resulting expression  for the  $\vec{k}=\vec{0}$  Fourier transform of 
the the pressure tensor :  
\begin{mathletters}
\label{sigma}    
\begin{equation}
    \sigma_{\alpha,\beta}=\sigma_{\alpha,\beta}^{K} + \sigma_{\alpha,\beta}^{d}
    +\sigma_{\alpha,\beta}^{f} \; ,
\end{equation}   
\begin{equation}
    \sigma_{\alpha,\beta}^{K}=m \sum_{i=1}^{N} v_{i,\alpha} 
    v_{i,\beta}
    \; ,
\end{equation}
\widetext
\begin{equation}
    \sigma_{\alpha,\beta}^{d}=-\frac{q^2}{2} \sum_{i\neq j} 
    \frac{r_{ij,\alpha} r_{ij,\beta}}{\|\vec{r}_{ij}\| } \; 
   \left[ \sum_{\epsilon=\pm 1}\frac{d}{dr}\left.
    \frac{\text{erfc}(\delta \|\vec{r}\| +\epsilon 
\alpha /2 \delta)\; \exp( \epsilon \alpha r)}{2r} \right |_{r=\| 
\vec{r}_{ij} \| } \right] \; ,
\end{equation}    
\begin{equation}
    \sigma_{\alpha,\beta}^{f}=\frac{2\pi q^2}{L^{3}} \sum_{i\neq j} 
    \sum_{ \|\vec{k}\|
\le k_0} 
    \frac{\exp\left(-\left(\vec{k}^{2}+\alpha^{2}\right)/4\delta^{2}\right)}
    {\vec{k}^{2} + \alpha^{2}}
    \exp(i\vec{k}\cdot \vec{r}_{ij})
  \left( \delta_{\alpha,\beta} - (\frac{1}{2\delta^{2}} 
  +\frac{2}{\vec{k}^{2}+\alpha^{2}})k_{\alpha}k_{\beta}  
  \right) \; .
\end{equation}    
\end{mathletters}   
\narrowtext

In our simulations we choose as unit of length the ionic radius $a$ 
and as unit of time $\tau=\sqrt{3} \omega_{p}^{-1}$ with  
$\omega_{p}^{2}=4\pi \rho q^{2}/m$. The calculations were performed in the 
microcanonical ensemble and the trajectories of each of the $N$ 
particles (and all its images) were computed by a time-symmetrical 
integer algorithm \cite{Verlet}. This algorithm  is symplectic and 
ensures an exact 
conservation of the total momentum of the system.
The time increment $\Delta t$
was chosen in such a way to ensure a 
good conservation of the energy (typically $\Delta t=10^{-2}\tau$ leads to
fluctuations  of $\sim 10^{-7}$ on the
average energy). In most of our simulations 
$N=500$, but smaller and larger systems were also considered in order to 
study finite size effects on the transport coefficients. 
Typically  $5 
\times 10^{5}$ time steps were generated after a careful 
equilibration of the system. Each run was divided
into statistically independent blocks of $\sim 5 \times 10^4$ time steps, i.e.
much larger than the correlation time.  
The reported errors on the autocorrelation functions and the transport
coefficients 
were obtained by a standard block analysis \cite{Til}; they correspond 
to one standard deviation. As an example of the precision which can be
obtained for sufficiently long calculations, we display  in Fig.\ \ref{fig1} 
the autocorrelation of
the energy current at $(\Gamma=10,\alpha^*=0.1)$. The integral of the
function over the time reaches a well defined plateau which allows for an
accurate determination of the thermal conductivity. The precision on
$\lambda$ and on the other transport coefficients is typically of the order
of $\sim 1 \%$ for most of the considered cases.

Since the thermodynamic states of the YOCP are characterized by two
parameters, a systematic determination of the transport coefficients in the
whole fluid phase requires an enormous amount of simulations.
 We present here only preliminary results for a few
thermodynamic states; extended results will be given in a forthcoming
publication \cite{SC1}. Our results are summarized in Tables\ I, II, III. We choose
the following units: $m \omega_p \rho a^2$ for the viscosities ($\eta =
m \omega_p \rho a^2 \eta^*$, $\xi = m \omega_p \rho a^2 \xi^*$) , $k \omega_p
\rho a^2$ for the thermal conductivity ($\lambda= k \omega_p
\rho a^2 \lambda^*$).

In order to check our method we have first considered the case $\alpha^*=0.01$
and compared our results with those of Bernu and Vieillefosse \cite{Bernu}
and of Donk\'o  {\em et al.}
\cite{Donko} for the OCP. The former authors have performed EMD 
simulations of the OCP and give estimates of ($\eta, \; \xi, \; \lambda$) for 
a few thermodynamic states,  while the 
latter provide extensive NEMD computations of $\eta$ and $\lambda$.
As far as the shear viscosity is concerned, all the results are in overall good
agreement except at low $\Gamma$'s. However, it must be stressed that,
in this regime,  Bernu and Vieillefosse have considered only relatively 
small systems of $N=128$ particles;  their 
results
for $\eta$ are hence  probably underestimated due to finite size effects. 
As seen from table\ I, the reduced bulk viscosity  $\xi^*$ is 
typically three orders of
magnitude smaller than $\eta^*$ which makes difficult its
precise determination and entails relatively important statistical errors.
Our estimates of  $\xi^*$ agree well with those of
Bernu {\em et al.} at large $\Gamma$'s but, as for the shear 
viscosity, differ significantly at low 
$\Gamma$'s.  Finally,  our results for the thermal
conductivity at $\alpha^*=0.01$ are in good
agreement with those of Bernu {\em et al.} (except for the lowest 
$\Gamma$'s)
but are systematically higher than those obtained by  Donk\'o  {\em et al.} in
their NEMD simulations \cite{Donko}.

The recent NEMD simulations of Sanbonmatsu and M. S. Murillo \cite{Murillo}
on the YOCP have been performed
only for large values of the screening parameter ( i.e. for 
$\alpha^*=1,2,3,4$). In this case, Ewald sums can probably be safely 
ignored, at least for sufficiently large systems.
  Our estimates of the shear vicosity 
  are compared with those of ref.\ \cite{Murillo} in table II for  a few
thermodynamic states. The disagreement between the two series of
simulations is patent, particularly for large values of $\alpha^*$ where the
results may differ by a factor of $\sim 4$. In order to clarify this 
point, we have focussed on the case of the large  $\alpha^*$'s.
In this regime  we actually
deal with a dilute gas of particles interacting via  short range 
potentials. Clearly, in this case, the transport coefficients can be computed
in the framework of the Chapman-Ensog (CE) theory\cite{Chapman}.
In the so-called first CE approximation, we have  $\xi^*=0$ and 
\begin{equation}
\eta_{CE} = \frac{5}{8} \frac{k T}{\Omega^{(2)}(2)} \; ,
\label{CEb}
\end{equation}
where $\Omega^{(2)}(2)$ is a standard collision integral 
\cite{Chapman}.
Note first that $\xi^*=0$ which is compatible with the low values of the
reported data and the 
steady decay  of $\xi^*$ with respect to $\alpha^*$ for a fixed $\Gamma$,
as seen from table II. 
Moreover, it can be shown that the expression\ (\ref{CEb}) of the CE shear viscosity 
of the YOCP can be rewritten as
\begin{equation}
\eta_{CE} = \frac{\alpha^{*2}}{\sqrt{\Gamma}} \; {\cal I}(\alpha^* \Gamma) \; ,
\label{CE}
\end{equation}
where ${\cal I}(\alpha^* \Gamma)$ is a triple integral that we have computed
numerically by
Monte Carlo integration methods. In Fig.\ 2 we display the EMD and CE 
shear viscosities as functions of $\alpha^*$ for 
$\Gamma=2, \;10, \; 50$. The agreement between our EMD results
and the predictions of the CE theory is obvious for sufficiently
large $\alpha^*$'s.
More precisely, the CE estimates seem to be accurate as soon as the 
coupling parameter $\Gamma
\exp(-\alpha^*) \lesssim 0.35$. The CE theory also enables the computation of
the thermal conductivity $\lambda_{CE}=5C_v\eta_{CE}/2$ and we found, as in the case of the shear
viscosity, a perfect agreement between our EMD
simulations and the CE theoretical predictions in the domain $\Gamma
\exp(-\alpha^*) \lesssim 0.35$. 

To summarize, our EMD results for the transport coefficients of the YOCP are in
good agreement with the available data on  the OCP in the limit
 $\alpha^* \to 0$  and
also in good agreement with the predictions of the CE theory for large
values of $\alpha^*$, as it should be, and in severe disagreement 
with the values reported in ref.\ \cite{Murillo}. 

We think that the standard approach used in this work 
to compute the transport coefficients  -i.e. EMD 
simulations plus Ewald sums - is  efficient and reliable for the two
following reasons : 
\begin{itemize}
\item the three transport coefficients $\eta$, $\xi$, and $\lambda$ can be
computed in a single run. By contrast, each transport coefficient 
requires a separate NEMD simulation.
\item thanks to Ewald sums the simulations can be undertaken for any value of
$\alpha^*$ and they require  a small number $N$ of particles. By 
contrast NEMD simulations seem to require larger system 
sizes which precludes the use of Ewald sums\cite{Murillo,Donko}. 
\end{itemize}

 We have indeed 
checked that finite size
effects on the transport coeeficients are small as long as $N \geq 256$. For
instance, for the state $(\Gamma=10,\alpha^*=1$), we found for the thermal
conductivity $\lambda^*=0.4138(41),\; 0.5556(54),\; 0.5397(69),\; 0.5372(56)$
 for $N=128,256,500,864$ respectively.
Therefore systems of $N\sim 500$ are sufficiently large to ensure a reliable
estimate of the transport coefficients. Some discrepencies between our 
results in the case $\alpha^{*}=0.01$ and those obtained by Bernu {\em et al.} 
for the OCP with systems made of $N=128$ particles probably
originate in finite size effects.

Finally we have also considered
small values of the screening parameter $\alpha^*$, i.e.
$0\leq \alpha^*\leq 1$.
In this case the use of Ewald sums cannot be avoided and some  
preliminary results are
diplayed in table III. Calculations are in progrees for other values 
of $(\Gamma,\alpha^{*})$ and many 
more results will be given together with a fit of all transport 
coefficients as functions of $(\Gamma,\alpha^{*})$ \cite{SC1}. 

We acknowledge  D. Gilles, J. Cl\'erouin, and D. Levesque for 
useful discussions and D. Levesque for providing us a MD code of the 
OCP easily transformed in a MD code for the YOCP. 

%%%%%%%%%%%%%%%%%%%%%%%%%%%%%%%%%%%%%%%%%%%%%%%%%%%%%%%%%%%%%%%%%%%%%%%%%%%%%%%%
%%%%%%%%%%%%%%%%%%%%%%%%%%%%%%%%%%%%%%%%%%%%%%%%%%%%%%%%%%%%%%%%%%%%%%%%%%%%%%%%

%%%%%%%%%%%%%%%%%%%%%%%%%%%%%%%%%%%%%%%%%%%%%%%%%%%%%%%%%%%%%%%%%%%%%%%%%%%
%%%%%%%%%%%%%%%%%%%%%%%%%%%%%%%%%%%%%%%%%%%%%%%%%%%%%%%%%%%%%%%%%
\begin{figure}
\caption{Solid curve : the autocorrelation function of the energy current 
$\langle \vec{J}_{e}(t) \cdot
\vec{J}_{e}(0) \rangle$  for ($\Gamma=10,\alpha^*=0.1$), dotted
 curve : cumulative sum.}
\label{fig1}
\end{figure}
%%%%%%%%%%%%%%%%%%%%%%%%%%%%%%%%%%%%%%%%%%%%%%%%%%%%%%%%%%%%%%%%%
\begin{figure}
\caption{Shear viscosity of the YOCP as a function of $\alpha^{*}$ for 
various $\Gamma$'s. Solid
curve : EMD results, Dashed curve : CE theory. }
\label{fig2}
\end{figure}
%%%%%%%%%%%%%%%%%%%%%%%%%%%%%%%%%%%%%%%%%%%%%%%%%%%%%%%%%%%%%%%%%%
\mediumtext
\begin{table}
\caption{Transport coefficients of the YOCP in the limit $\alpha^* \to 0$.
The numbers in brackets denote the accuracy of the last digits. }
\begin{tabular}{ccccccccc}
&\multicolumn{3}{c}{$\eta$}&\multicolumn{2}{c}{$\xi \times 10^{-3}$}&\multicolumn{3}{c}{$\lambda$}\\
\cline{2-4} \cline{5-6} \cline{7-9}
$\Gamma$&YOCP\tablenotemark[1]&OCP\tablenotemark[2]&OCP\tablenotemark[3]
        &YOCP\tablenotemark[1]&OCP\tablenotemark[2]
        &YOCP\tablenotemark[1]&OCP\tablenotemark[2]&OCP\tablenotemark[3]\\
$1     $&$ 1.16(5)  $&$ 1.04 (21) $&           &$ 4.72(36) $&$2.6  (6)$&$ 4.24(29)  $&$ 2.9(6)   $&$\sim 2.2 $\\
$2     $&$ 0.527(7) $&             &$\sim 0.5 $&$ 3.02(5)  $&$        $&$ 1.862(16) $&$          $&$\sim 1.2 $\\
$10    $&$ 0.112(1) $& $0.085 (17)$&$\sim 0.1 $&$ 1.753(24)$&$1.8 (5) $&$ 0.5586(70)$&$ 0.66 (16)$&$\sim 0.40$\\
$100   $&$0.1874(20)$&$ 0.18(3)   $&$\sim 0.18$&$ 0.394(7) $&$0.21(6) $&$ 0.843(11) $&$ 0.88 (17)$&$\sim 0.72$\\
\end{tabular}
\tablenotetext[1]{EMD results at $\alpha^*=0.01$}
\tablenotetext[2]{EMD results of Bernu and Vieillefosse
 Ref.\ \cite{Bernu} for the OCP.}
\tablenotetext[3]{NEMD results of Donk\'o {\em et al.} Ref.\ \cite{Donko} for the OCP.}
\end{table}

%%%%%%%%%%%%%%%%%%%%%%%%%%%%%%%%%%%%%%%%%%%%%%%%%%%%%%%%%%%%%%%%%%%%
%\widetext
%\begin{table}
%\caption
%{Transport coefficients of the YOCP for few thermodynamic 
%states. For each  coefficient,  first column : present EMD results, 
%second column : Chapmann-Enskog prediction ($\xi_{CE}=0$ not reported)
%, third column (only for $\eta^{*}$) NEMD estimates of 
%ref.\ [5]. The numbers in brackets denote the accuracy of the 
%last digits.} 
%\begin{tabular}{ccccccccccccc}
%&\multicolumn{6}{c}{$\Gamma=2$}&\multicolumn{6}{c}{$\Gamma=10$}\\
%\cline{2-7} \cline{8-13}
%$\alpha$&\multicolumn{3}{c}{$\eta$}&\multicolumn{2}{c}{$\lambda$}&$\xi$
%        &\multicolumn{3}{c}{$\eta$}&\multicolumn{2}{c}{$\lambda$}&$\xi$\\
%\cline{2-4} \cline{5-6} \cline{7-7} \cline{8-10} \cline{11-12} \cline{13-13}
%$1$&$0.496(12)$&$0.439(64) $&$0.2340$&$2.42(12) $&$1.65(24) $&$0.834(48)    $
%   &$0.112(3) $&$0.047(5)  $&$0.0526$&$0.570(18)$&$0.176(18)$&$1.282(48)    $\\
%$2$&$0.991(24)$&$0.826(48) $&$0.2646$&$2.89(17) $&$3.09(18) $&$0.756(14)    $ 
%   &$0.145(3) $&$0.117(10) $&$0.0521$&$0.644(17)$&$0.438(39)$&$1.205(48)(14)$\\
%$3$&$1.282(36)$&$1.367(94) $&$0.4760$&$5.36(29) $&$5.13(35) $&$0.694(12)    $ 
%   &$0.198(3) $&$0.193(10) $&$0.0693$&$0.841(18)$&$0.726(40)$&$1.426(12)    $\\
%$4$&$1.935(36)$&$2.055(152)$&$0.5496$&$7.18(23) $&$7.71(57) $&$0.447(5)     $
%   &$0.306(4) $&$0.288(19) $&$0.0870$&$1.239(23)$&$1.08(7)  $&$1.255(9)     $\\
%\end{tabular}
%\end{table}

\begin{table}
\caption
{Transport coefficients of the YOCP for few thermodynamic 
states. For each  coefficient,  first column : present EMD results, 
second column : Chapmann-Enskog prediction ($\xi_{CE}=0$ not reported)
, third column (only for $\eta^{*}$) NEMD estimates of 
ref.\ [5]. The numbers in brackets denote the accuracy of the 
last digits.} 
\begin{tabular}{ccccccc}
&\multicolumn{6}{c}{$\Gamma=2$}\\
\cline{2-7} 
$\alpha$&\multicolumn{3}{c}{$\eta$}&\multicolumn{2}{c}{$\lambda$}&$\xi \times 10^{-3}$
\\
\cline{2-4} \cline{5-6} \cline{7-7} 
$1$&$0.496(12)$&$0.439(64) $&$0.2340$&$2.42(12) $&$1.65(24) $&$0.834(48)    $\\
$2$&$0.991(24)$&$0.826(48) $&$0.2646$&$2.89(17) $&$3.09(18) $&$0.756(14)    $\\
$3$&$1.282(36)$&$1.367(94) $&$0.4760$&$5.36(29) $&$5.13(35) $&$0.694(12)    $\\
$4$&$1.935(36)$&$2.055(152)$&$0.5496$&$7.18(23) $&$7.71(57) $&$0.447(5)     $\\
\end{tabular}
\begin{tabular}{ccccccc}
&\multicolumn{6}{c}{$\Gamma=10$}\\
\cline{2-7} 
$\alpha$&\multicolumn{3}{c}{$\eta$}&\multicolumn{2}{c}{$\lambda$}&$\xi \times 10^{-3}$
\\
\cline{2-4} \cline{5-6} \cline{7-7} 
$1$&$0.112(3) $&$0.047(5)  $&$0.0526$&$0.570(18)$&$0.176(18)$&$1.282(48)    $\\
$2$&$0.145(3) $&$0.117(10) $&$0.0521$&$0.644(17)$&$0.438(39)$&$1.205(48)(14)$\\
$3$&$0.198(3) $&$0.193(10) $&$0.0693$&$0.841(18)$&$0.726(40)$&$1.426(12)    $\\
$4$&$0.306(4) $&$0.288(19) $&$0.0870$&$1.239(23)$&$1.08(7)  $&$1.255(9)     $\\
\end{tabular}
\end{table}

\begin{table}
\caption{Transport coefficients of the YOCP for small values of the 
screening parameter $\alpha$. The numbers in brackets denote 
the accuracy of the last digits. }
\begin{tabular}{cccccccccc} 
&\multicolumn{3}{c}{$\eta$}&\multicolumn{3}{c}{$\xi \times 10^{-3}$}&\multicolumn{3}{c}{$\lambda$}\\
\cline{2-4} \cline{5-7} \cline{8-10}
$\alpha$ &$\Gamma=2$&$\Gamma=10$&$\Gamma=50$
         &$\Gamma=2$&$\Gamma=10$&$\Gamma=50$
         &$\Gamma=2$&$\Gamma=10$&$\Gamma=50$\\
 $0.2$   &$0.513(5)$&$0.1054(12)$&$0.1102(7)$&$2.78(5)$&$1.287(24)$&$0.534(7)$&$1.716(11)$&$0.55(1)$&$0.641(1)$ \\
 $0.4$   &$0.464(5)$&$0.1033(12)$&$0.1069(7)$&$1.439(24)$&$1.238(24)$&$0.451(5)$&$1.96(2)$&$0.55(1)$&$0.704(1)$ \\
 $0.6$   &$0.513(5)$&$0.1093(17)$&$0.1016(6)$&$0.967(12)$&$0.914(17)$&$0.3906(85)$&$1.99(2)$&$0.518(9)$&$0.560(1)$ \\
 $0.8$   &$0.525(5)$&$0.1028(15)$&$0.0937(6)$&$0.851(5)$&$0.788(19)$&$0.372(5)$&$2.36(2)$&$0.492(12)$&$0.592(1)$ \\
\end{tabular}

\end{table}

%%%%%%%%%%%%%%%%%%%%%%%%%%%%%%%%%%%%%%%%%%%%%%%%%%%%%%%%%%%%%%%%%%%%%%%%%%%%%%

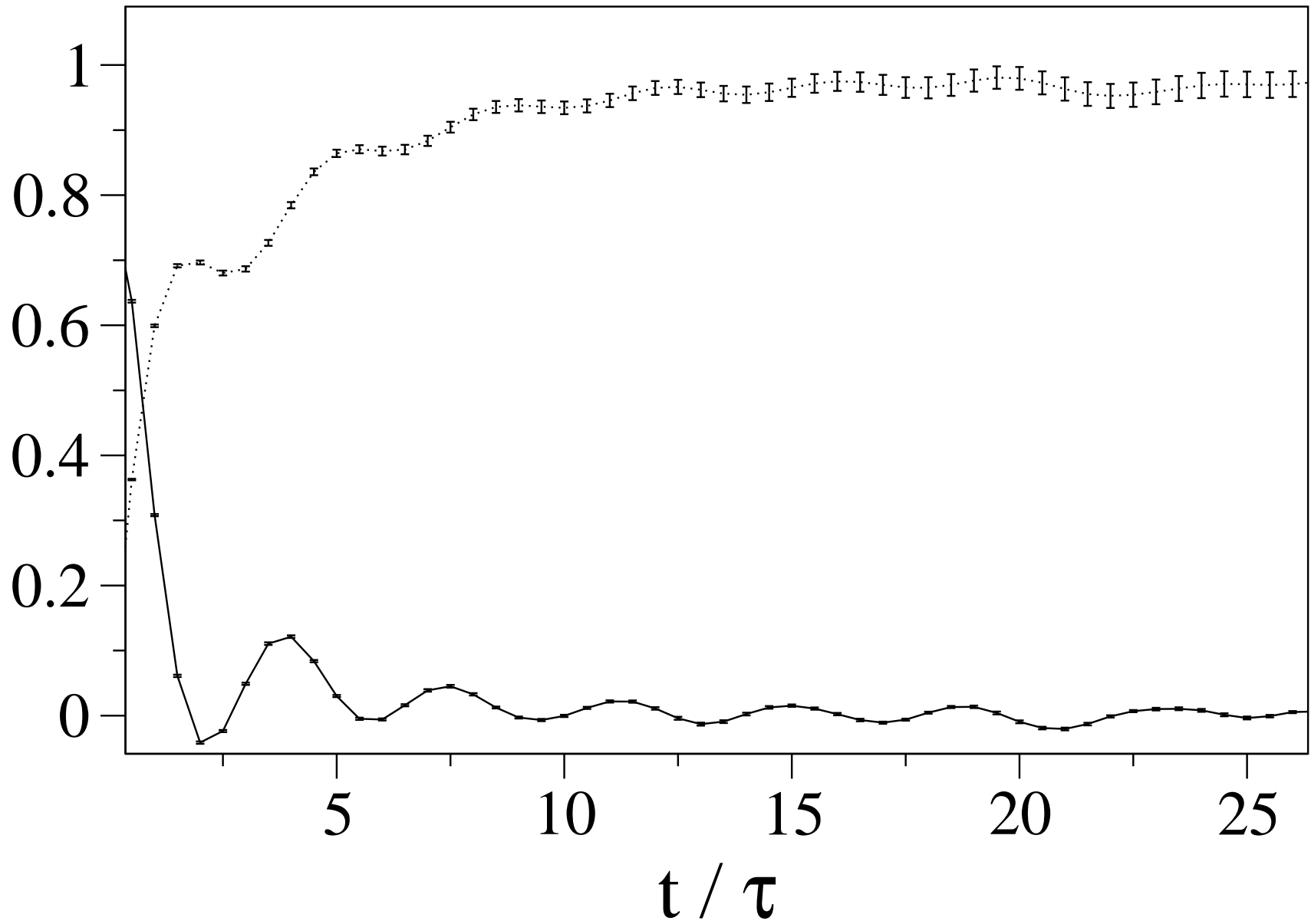
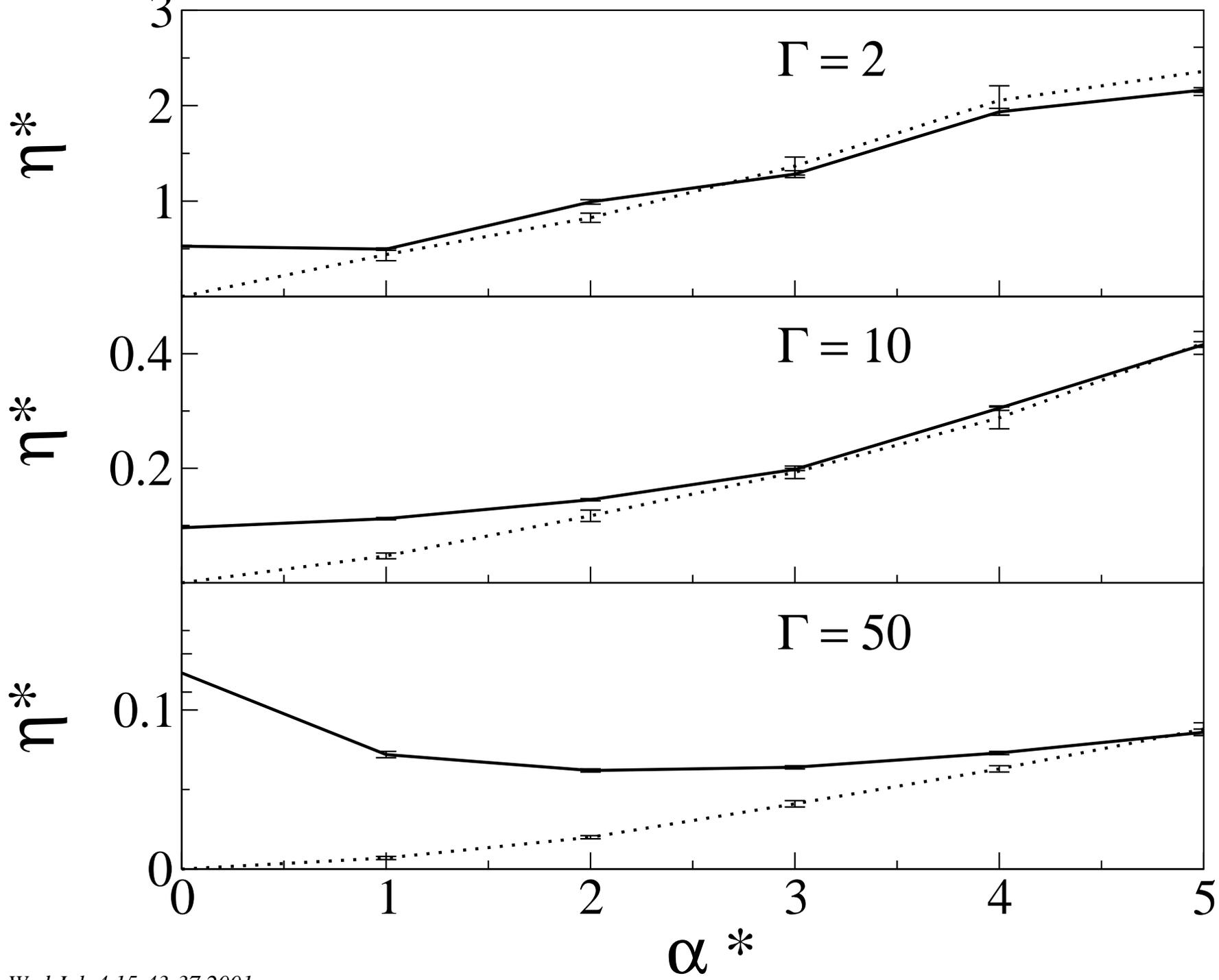
\end{document}